\documentclass{PoS}

\usepackage{epsfig}

\title{Comparison of differential elastic cross sections in $pp$ and $p\bar{p}$ collisions as evidence of the existence of the
colourless $C$-odd three-gluon state}

\ShortTitle{Elastic cross sections in $pp$ and $p\bar{p}$ collisions}

\author{\speaker{Christophe Royon} \\  
        The University of Kansas, Lawrence, USA\\
        E-mail: \email{christophe.royon@ku.edu} \\
        On behalf of the D0 and TOTEM Collaborations}

\abstract{We analyze the differences between the $pp$ and $p\bar{p}$ differential elastic cross section measurements by the
D0 and TOTEM Collaborations at the Tevatron, Fermilab, and the LHC, CERN that lead to a significance larger than 3$\sigma$ of the existence of the
colourless $C$-odd three-gluon state, the odderon.}

\FullConference{40th International Conference on High Energy physics - ICHEP2020\\
  July 28 - August 6, 2020\\
  Prague, Czech Republic (virtual meeting)}

\begin{document}

Looking for evidence of the existence of the odderon exchange has been the subject of many experiments  since its original model in Ref.~\cite{nicolescu}. It remained elusive until now. The odderon is defined as a singularity in the complex plane which contributes to the odd crossing amplitude. In terms of QCD, this leads 
to the contributions of 2, 3... gluon exchanges~\cite{bkp}. The colorless $C$-odd 3-gluon state (odderon)  exchange predicts differences in elastic $d\sigma/dt$ for $pp$ and $p\bar{p}$ interactions since it corresponds to different amplitudes and interferences. This is why the TOTEM and D0 Collaborations decided to collaborate in order to analyze the potential differences between elastic $pp$ and $p \bar{p}$ interactions, that they measured respectively at 2.76, 7., 8., 13. TeV~\cite{totemelastic} and 1.96 TeV~\cite{d0elastic}.  The difficulty of this study is that the measurements were not performed at the same center-of-mass energies and one needs to extrapolate for instance the $pp$ TOTEM measurements to the Tevatron energy~\cite{d0note}.

\section{Elastic scattering at the Tevatron and the LHC and the bump over dip ratio}
\begin{figure}
\begin{center}
\epsfig{figure=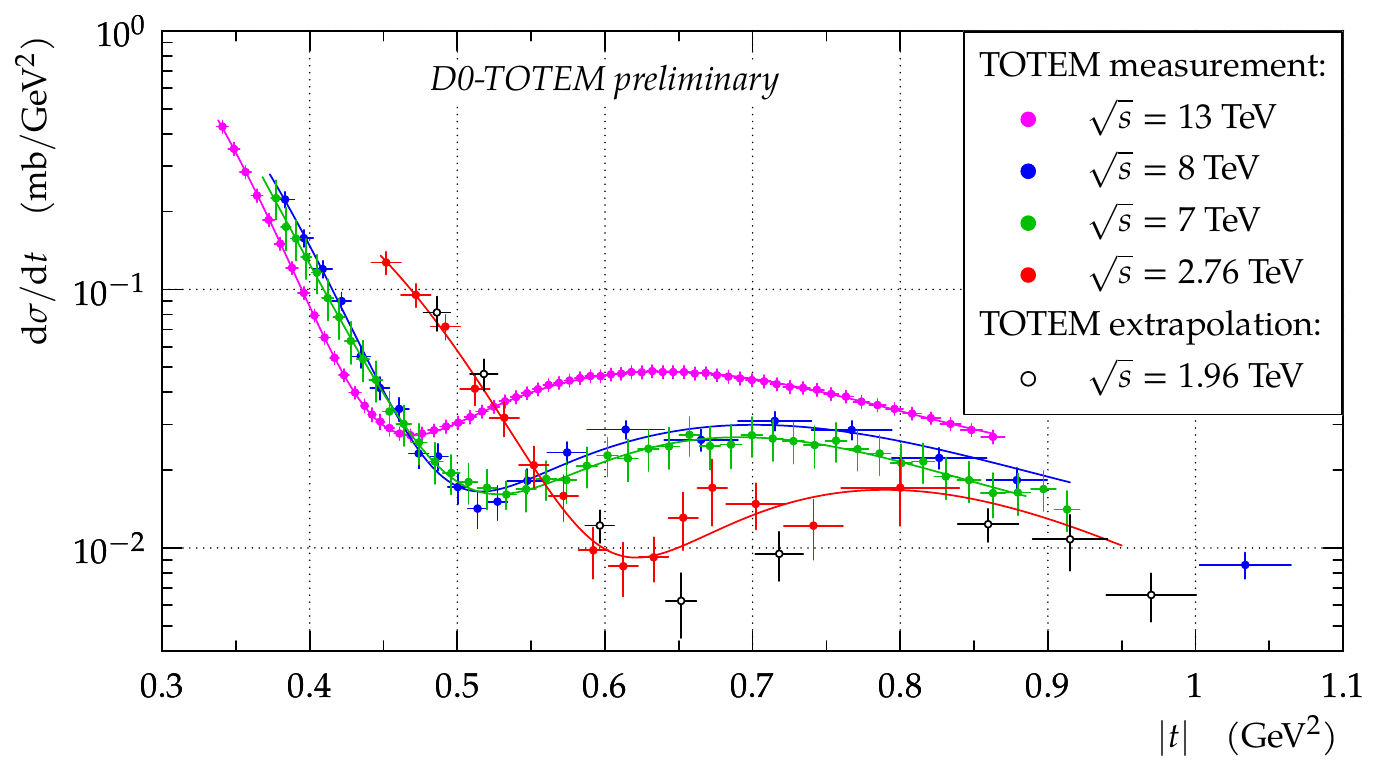,height=0.35\textwidth}
\caption{Elastic $pp$  $d \sigma/dt$ cross section measurements from the TOTEM  collaboration and extrapolation at 1.96 TeV.}
\label{fig1}
\end{center}
\end{figure}

The elastic $pp \rightarrow pp$ and $p\bar{p} \rightarrow p\bar{p}$ cross sections were measured respectively at the LHC at 2.76, 7, 8 and 13 TeV with unprecedented precision by the TOTEM Collaboration~\cite{totemelastic} as illustrated in Fig.~\ref{fig1} and at 1.96 TeV at the Tevatron by the D0 Collaboration~\cite{d0elastic}. These reactions correspond to the  exchange of momentum
between the two protons which remain intact. It is thus possible to
measure the intact $p$ or $\bar{p}$ scattered close to the beam using roman pots installed both by D0 and TOTEM Collaborations and located respectively at about 56 and 220 m from the D0 and CMS interaction points. The idea s to move the detectors inside the roman pots as close to the beam as possible (typically a few sigmas) in order to measure the $p$ and $\bar{p}$ scattered at very small angles. 
From counting the number of events as a function of $|t|$, the squared quadri-momentum transferred 
at the $p$ or $\bar{p}$ vertex measured by tracking the protons, it is possible to compute the elastic differential cross section $d\sigma/dt$.

The strategy is then to compare extrapolated $pp$ and measured $p \bar{p}$  elastic $d \sigma/dt$ at 1.96 TeV since there is no $pp$ and $p \bar{p}$ data at the same energy~\footnote{Running the LHC at the Tevatron energy would not help since the $t$ acceptance where the elastic cross section could be measured would be very limited, and would be outside the dip and bump region.}.  We will  identify characteristic features of TOTEM elastic $pp$ cross sections and extrapolate them at 1.96 TeV. In Fig.~\ref{fig1}, we notice that all TOTEM $d \sigma/dt$ measurements show the same features, namely the presence of a dip and a bump in data, whereas D0 data do not show this feature.

\begin{figure}
\begin{center}
\epsfig{figure=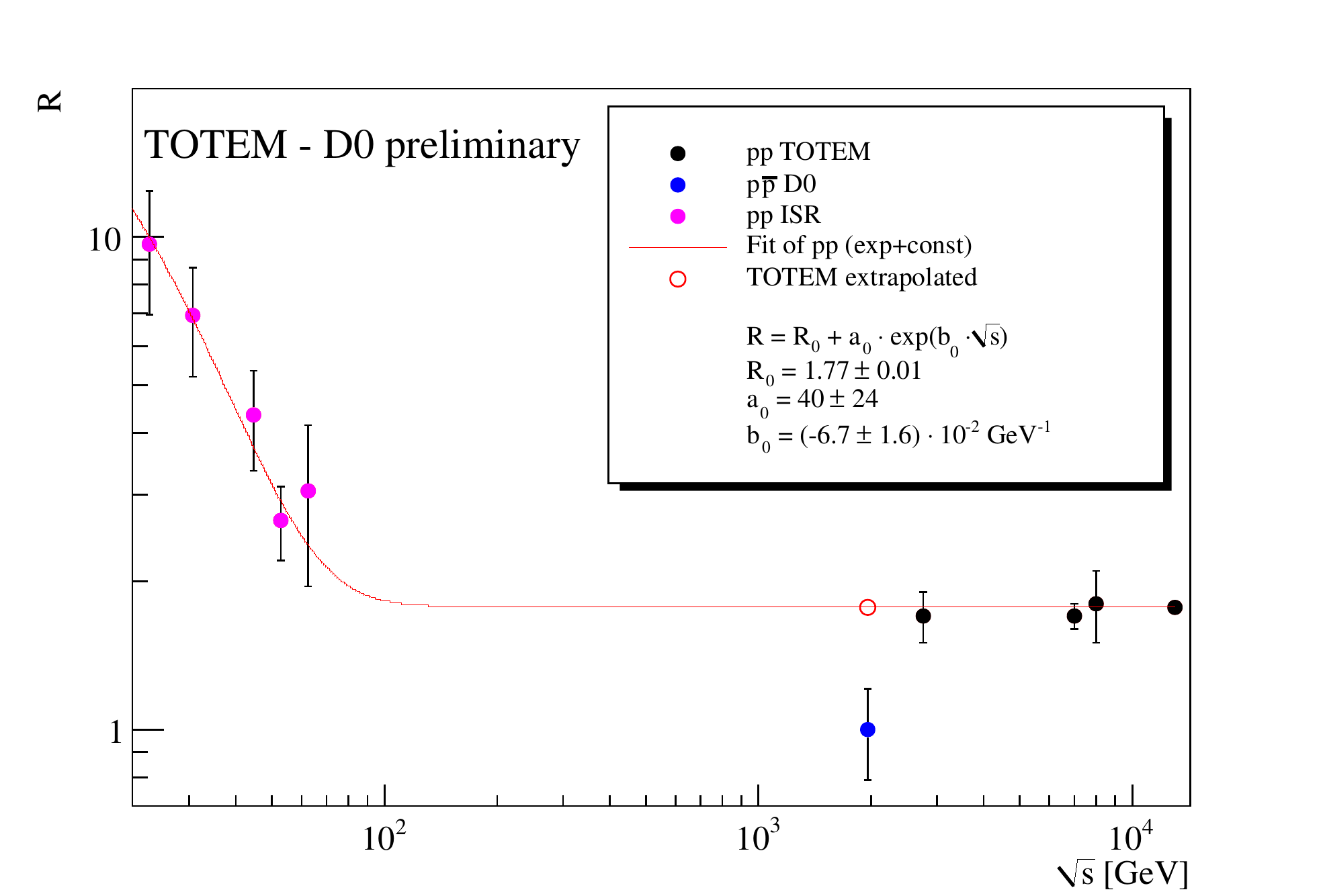,height=0.40\textwidth}
\caption{Bump over dip ratio for $pp$ intractions at the ISR and LHC energies.}
\label{fig2}
\end{center}
\end{figure}

The bump over dip ratio $R$ has been measured for $pp$ interactions at ISR and LHC energies as shown in Fig.~\ref{fig2}.
It decreases as a function of $\sqrt{s}$ up to $\sim$ 100 GeV and is flat above that energy.
On the contrary, D0 $p \bar{p}$ data show a ratio of 1 given the fact that neither a bump nor a dip is observed in data within uncertainties.
This leads to a difference by more than 3$\sigma$  between $pp$ and $p \bar{p}$ elastic data for the $R$ ratio, assuming  a flat behavior above $\sqrt{s} = 100 GeV$. In the following, we will consider a more elaborate method to compare $pp$ and $p \bar{p}$ elastic data.

\section{Comparison between elastic cross section measurements in $pp$ and $p \bar{p}$ interactions} 

\begin{figure}
\begin{center}
\epsfig{figure=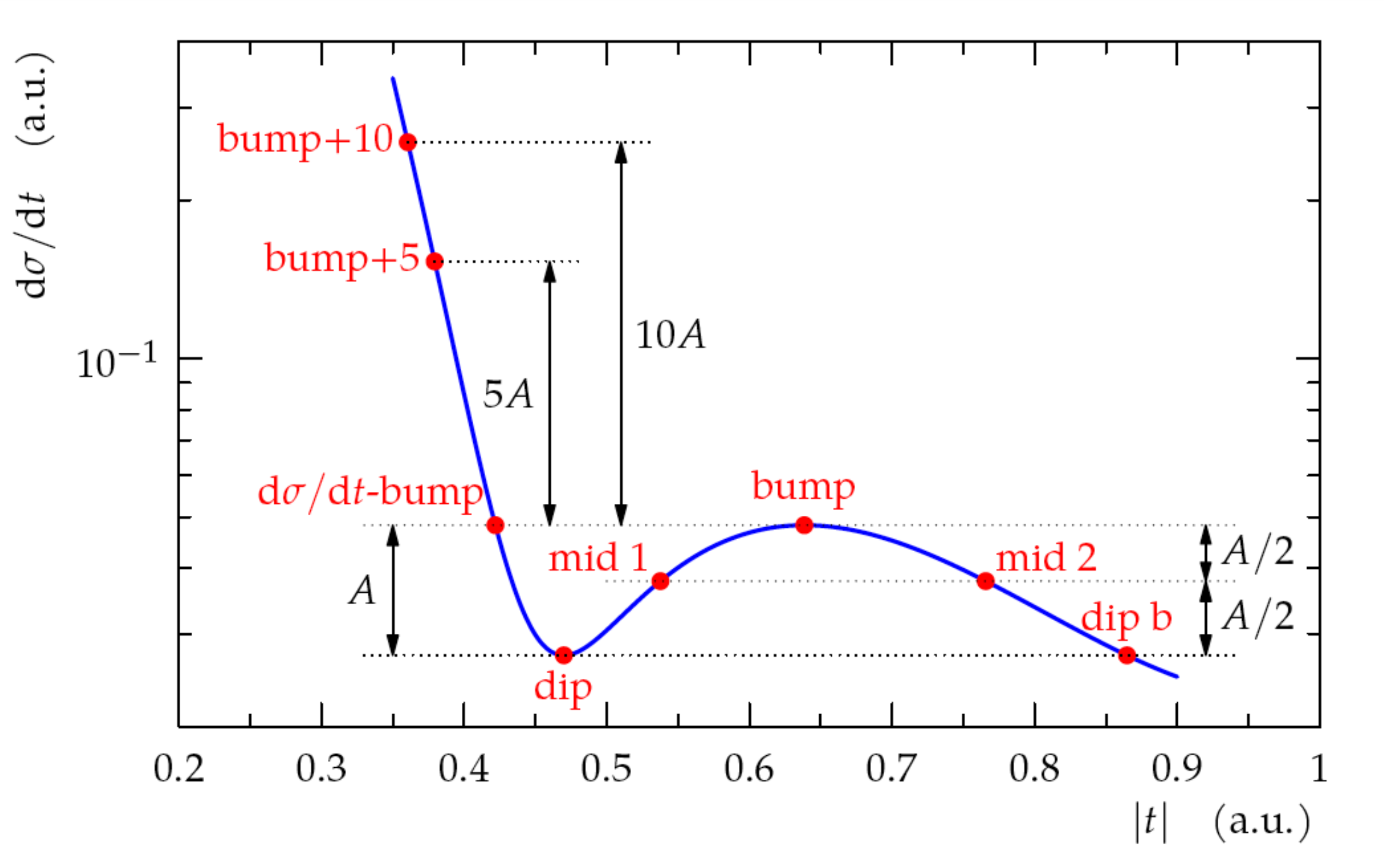,height=0.35\textwidth}
\caption{Definition of the eight characteristic points for the elastic $pp$ $d\sigma/dt$ cross sections.}
\label{fig3}
\end{center}
\end{figure}

\begin{figure}
\begin{center}
\epsfig{figure=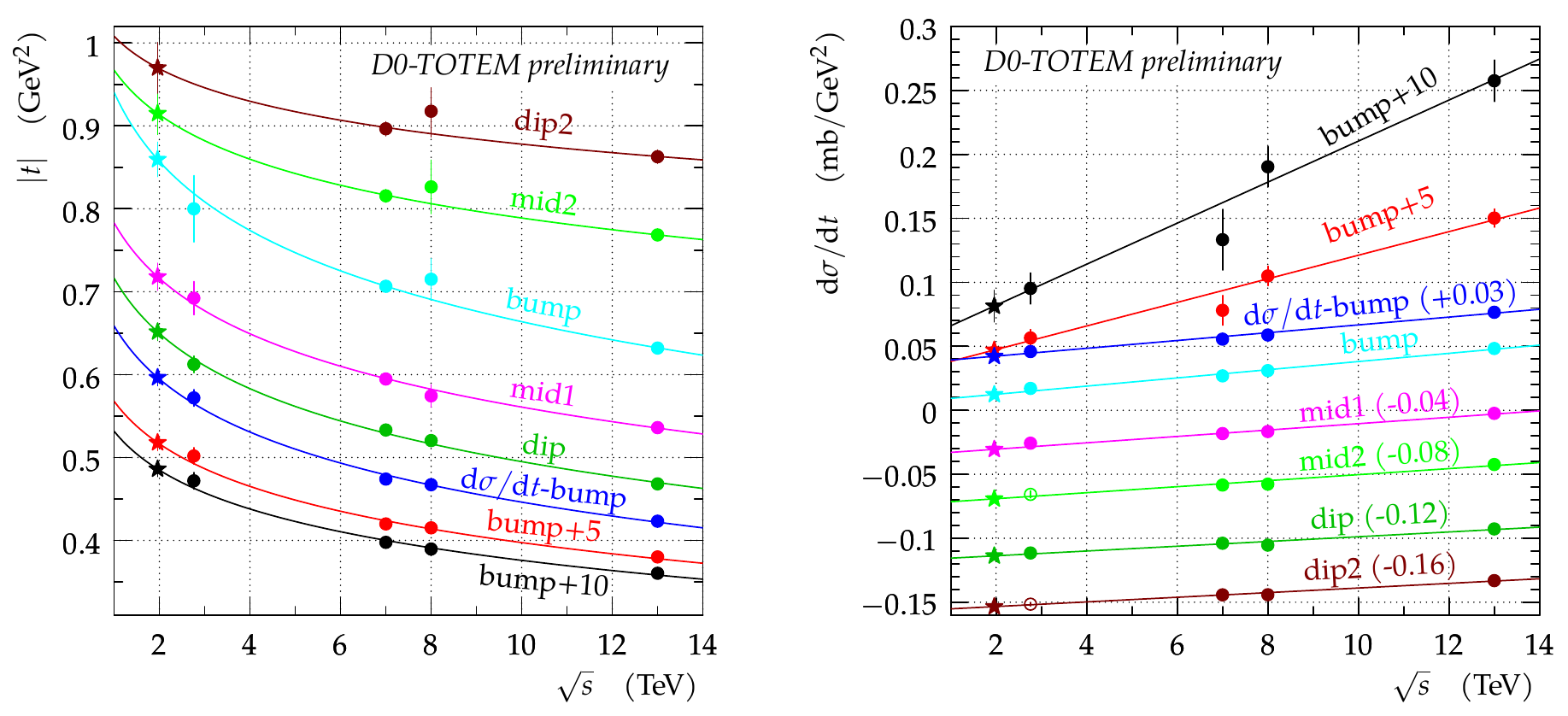,height=0.45\textwidth}
\caption{Variation of $t$ and $d \sigma/dt$ values for the eight characteristic points. The $d \sigma/dt$ values are shifted by the numbers indicated in parenthesis to distinguish between the different curves.}
\label{fig4}
\end{center}
\end{figure}

In this section, we will describe a method to extrapolate the TOTEM elastic measurements to Tevatron energy.
The first step is to define 8 characteristic points of elastic $pp$ $d \sigma/dt$ cross sections such as the dip, bump, etc, that represent the features of elastic $pp$ interactions as defined in Fig.~\ref{fig3}. We then
determine how the values of $|t|$ and $d\sigma/dt$ of these eight characteristic points vary as a function of $\sqrt{s}$ of the TOTEM measurements (2.76, 7, 8 and 13 TeV) in order to predict their values at the Tevatron energy, 1.96 TeV, as shown in Fig.~\ref{fig4}.  In order not to be model
dependent, we
use directly the published data points that are closest to those characteristic  points.
The $\sqrt{s}$ dependence of the characteristic points is then fitted to the following simple forms~\footnote{We also tried alternate parametrizations leading to similar results.}
\begin{eqnarray}
|t| = a \log (\sqrt{s}) + b  \label{eqn2}, 
~~~~~~~ (d\sigma/dt) = c \sqrt{s} + d \label{eqn3} \nonumber 
\end{eqnarray}
This leads to very good $\chi^2$ per dof, better than 1 for all fits.
By extrapolating the fit results to 1.96 TeV, we obtain predictions on $|t|$ and $d\sigma/dt$ at 1.96 TeV for each characteristic point. It is striking (and may be related to deep physics reasons) that the same formulae in $\sqrt{s}$ describes the $t$ and $d \sigma/dt$ values for all characteristic points.

We now determined the predicted differential elastic $pp$ cross section at 1.96 TeV for all $t$ values of the characteristic points and 
the last step is to predict the $pp$ elastic cross sections at the same $t$ values as measured by D0 in order to make a direct comparison.
For this sake, we fit the reference points extrapolated to 1.96 TeV from TOTEM measurements  using a double exponential fit ($\chi^2=0.63$ per dof)
\begin{eqnarray}
h(t) = a_1 e^{-b_1 |t|^2 - c_1|t|}  
+ d_1 e^{-f_1 |t|^3 - g_1 |t|^2 -h_1 |t|}. \nonumber
\end{eqnarray}
This function is chosen for fitting purposes only and  is justified by the fact that such a formula leads also to a good description of all TOTEM data in the dip and bump region at 2.76, 7, 8 and 13 TeV. 
The two exponential terms cross around the dip, one rapidly falling and becoming 
negligible in the high $t$-range whereas the other term rises off the dip.

\begin{figure}
\begin{center}
\epsfig{figure=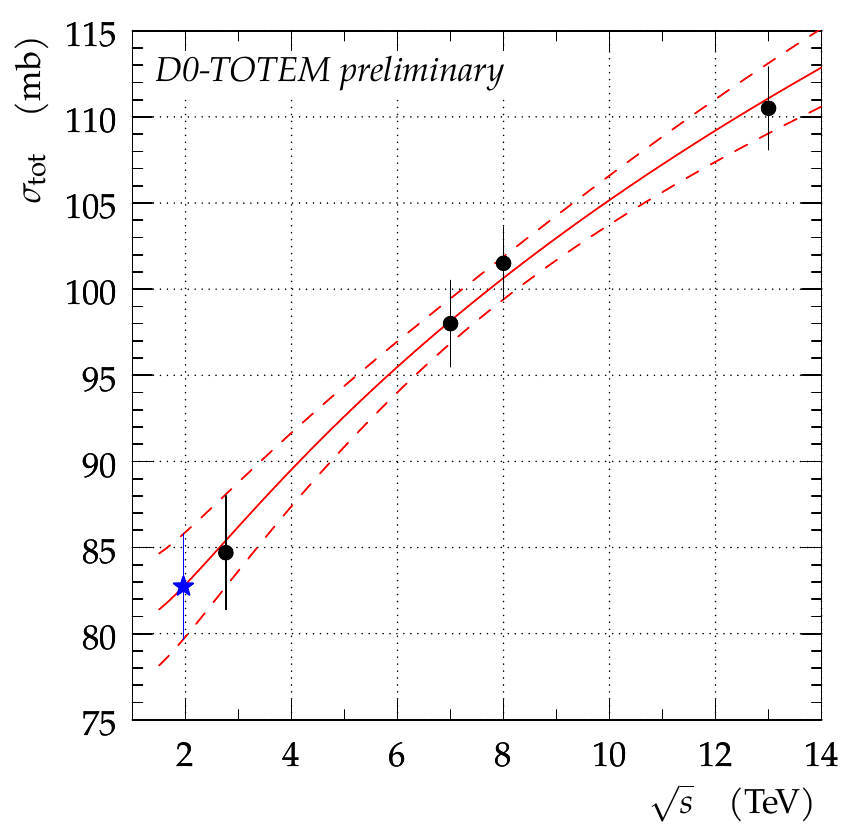,height=0.35\textwidth}
\caption{Extrapolated TOTEM data: total $pp$ cross section at 1.96 TeV.}
\label{fig5}
\end{center}
\end{figure}

We finally adjust the TOTEM extrapolated data following two constraints, the requirement that the $pp$ and $p \bar{p}$ optical points (OP), defined by $d\sigma/dt(t = 0)$, are
the same and that the logarithmic slopes of the cross sections are not modified in the normalization adjustment. 
OP cross sections are expected
to be equal if there are only $C$-even exchanges.
We first predict the $pp$ total  cross section from a fit to TOTEM data ($\chi^2=$0.27 for 2 dof) as shown in Fig.~\ref{fig5}
\begin{eqnarray}
\sigma_{tot} = a_2 \log^2 \sqrt{s} +b_2  \nonumber
\end{eqnarray}
which leads to an estimate of the $pp$ total cross section at 1.96 TeV of $\sigma_{tot}=$ 82.74 $\pm$ 3.06 mb.

Then, we can adjust the 1.96 TeV $d\sigma/dt(t=0)$ from extrapolating TOTEM data to the D0 measurement.
From the TOTEM $pp$ $\sigma_{tot}$, we obtain $d\sigma/dt(t=0)$ using the optical theorem
\begin{eqnarray}
\sigma_{tot}^2 = \frac{16 \pi }{1+\rho^2}  \left( \frac{d \sigma}{dt} \right)_{t=0}. \nonumber
\end{eqnarray}
Assuming $\rho=0.145$, the ratio of the imaginary and the real part
of the elastic nuclear amplitude, as taken from the COMPETE extrapolation~\cite{compete},
we obtain  $d\sigma/dt (t=0)=$ 357.1 $ \pm$ 26.4 mb/GeV$^2$ for the TOTEM data at the OP.
D0 measured the optical point of $d \sigma/dt$ at small $t$ to be 340.8 mb/GeV$^2$ and we
rescale the TOTEM data  by 0.954 $\pm$ 0.074~\footnote{We do
not claim that we performed a measurement of $d\sigma/dt$ at
the OP since it would require additional measurements  closer to $t = 0$, but
we use the two data sets simply in order to obtain
a common and somewhat arbitrary normalization point.}.

\begin{figure}
\begin{center}
\epsfig{figure=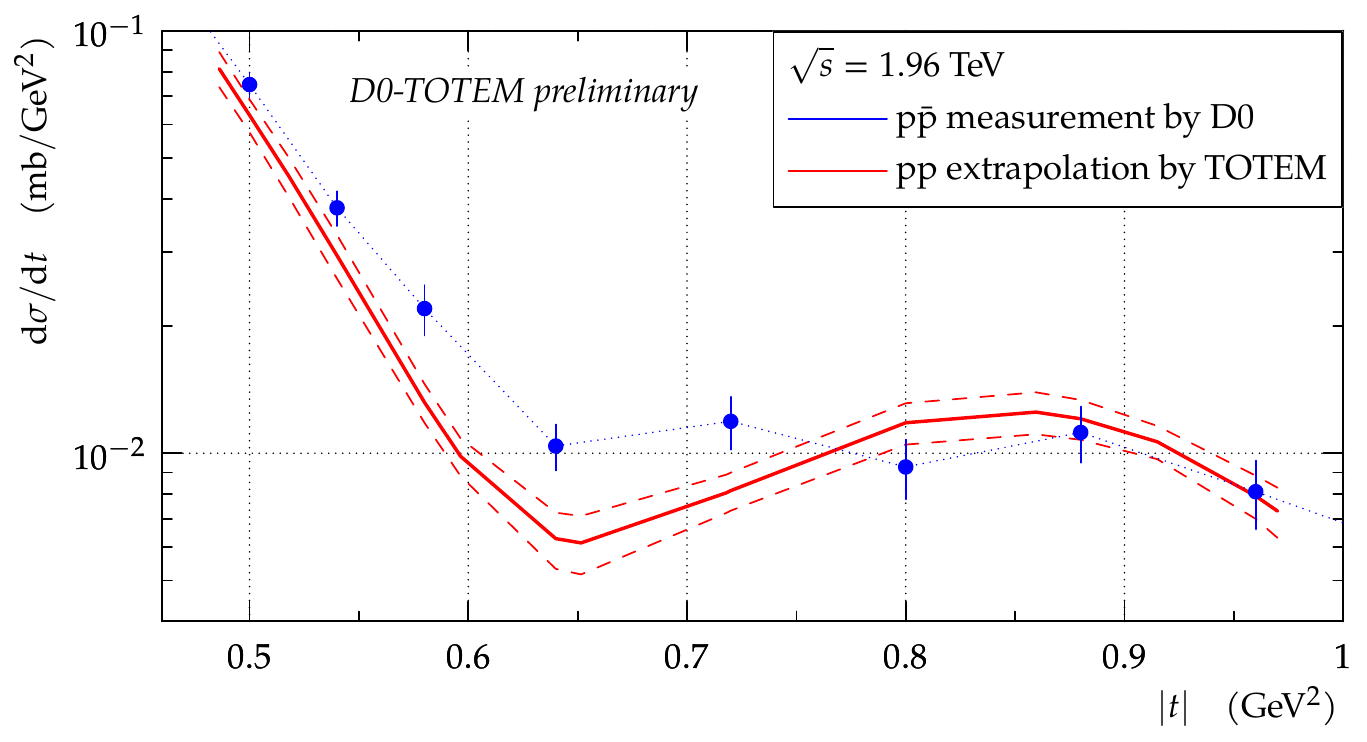,height=0.35\textwidth}
\caption{Comparison between elastic $p \bar{p}$ D0 measurement and $pp$ TOTEM extrapolation.}
\label{fig6}
\end{center}
\end{figure}

Two steps are then needed to compare quantitatively D0 and TOTEM extrapolated data: we consider the differences in shape and also in normalization.
We perform a Kolmogorov-Smirnov (KS) test sensitive to the difference in shape, and 
the  supremum value 
of the KS comparison is 0.0791 which leads to a  probability of
1.00\% for $pp$ and $p \bar{p}$ cross sections to be of the same shape, and a
corresponding significance of 2.33$\sigma$. 
In a second step, we compare the normalization. 
Comparing the $pp$ and the $p\bar{p}$
integrals of $d\sigma/dt$ in the same $|t|$-range of the measurement leads to a 2.11$\sigma$ effect, corresponding to a $p$-value of 1.73\%.
The combined significance using the Stouffer method~\cite{stouffer} leads to a  significance larger than  3.00$\sigma$ or to the evidence that the colorless $C$-odd three gluon state i.e. the odderon
is needed to explain elastic scattering at high energies. The comparison between the D0 measurement and the TOTEM extrapolated data is shown in Fig.~\ref{fig6}.

To conclude, we performed a 
detailed comparison between elastic $p\bar{p}$ (1.96 TeV from D0) and $pp$ data (2.76, 7, 8, 13 TeV from TOTEM). $pp$ and $p\bar{p}$ cross sections differ
with a significance over 3$\sigma$ in a model-independent way and thus provides evidence that the colorless $C$-odd three gluon state i.e. the odderon
is needed to explain elastic scattering at high energies. Combining this evidence with the independent $\rho$ measurement from TOTEM~\cite{rho}  definitely leads to a discovery of the odderon by more than 5$\sigma$~\cite{d0note}. 

\end{document}